# Fingerprints of the Cosmic Ray Driven Mechanism of the Ozone Hole

Qing-Bin Lu*

Department of Physics and Astronomy and Departments of Biology and Chemistry, University of Waterloo, 200 University Avenue West, Waterloo, Ontario, Canada

*Corresponding author (Email: qblu@uwaterloo.ca)

**ABSTRACT**
There is long research interest in electron-induced reactions of halogenated molecules. It has been two decades since the cosmic-ray (CR) driven electron-induced reaction (CRE) mechanism for the ozone hole formation was proposed. The derived CRE equation with stratospheric equivalent chlorine level and CR intensity as only two variables has well reproduced the observed data of stratospheric $O_3$ and temperatures over the past 40 years. The CRE predictions of 11-year cyclic variations of the Antarctic $O_3$ hole and associated stratospheric cooling have also been well confirmed. Measured altitude profiles of ozone and temperatures in Antarctic ozone holes provide convincing fingerprints of the CRE mechanism. A quantitative estimate indicates that the CRE-produced Cl atoms could completely deplete or even over-kill ozone in the CR-peak polar stratospheric region, consistent with observed altitude profiles of severest Antarctic ozone holes. After removing the natural CR effect, the hidden recovery in the Antarctic $O_3$ hole since ~1995 is clearly discovered, while the recovery of $O_3$ loss at mid-latitudes is being delayed by ≥10 years. These results have provided strong evidence of the CRE mechanism. If the CR intensity keeps the current rising trend, the Antarctic $O_3$ hole will return to the 1980 level by ~2060, while the returning of the $O_3$ layer at mid-latitudes to the 1980 level will largely be delayed or will not even occur by the end of this century. The results strongly indicate that the CRE mechanism must be considered as a key factor in evaluating the $O_3$ hole.

**INTRODUCTION**

In physics and chemistry, there is also a long research interest in electron-induced reactions of halogen-containing molecules.[1-5] Particularly for chlorofluorocarbons (CFCs, the major ozone-depleting substances), the dissociative electron attachment (DEA) of gaseous CFCs to low-energy free electrons near zero eV is highly effective, with measured DEA cross sections ~$10^4$ times the photodissociation cross sections of CFCs.[2,3] Peyerimhoff et al.[4,5] made the first suggestion that the DEA of CFCs must be seen in competition to the photodissociation process and must be considered as a factor in evaluating stratospheric $O_3$ depletion. However, the gaseous DEA process was thought to be insignificant for stratospheric CFCs,[6] though the understanding of stratospheric negative-ion chemistry was rather speculative.[7,8] The extremely effective dissociative electron transfer (DET) reaction of CFCs adsorbed on ice surfaces, which involved the prehydrated electron ($e_{pre}^-$) trapped in ice, was then surprisingly discovered by Lu, Madey and Sanche[9-13]. The DET cross sections were measured to be $(1.0-1.3)\times10^{-14}$ and $\sim 8.9\times10^{-14}$ cm$^2$ for $CF_2Cl_2$ and $CFCl_3$ on $H_2O$ ice respectively, $\sim 6.0\times10^{-12}$ cm$^2$ for $CF_2Cl_2$ on $NH_3$ ice, and $\sim 4.0\times10^{-15}$ cm$^2$ for HCl on $H_2O$ ice, which are $10^6$–$10^8$ times their photodissociation cross sections.[9-13] The highly effectively DET reaction of CFCs was later confirmed by other groups using real-time femtosecond laser spectroscopic methods.[14,15] Particularly, Wolf and co-workers[15,16] measured a large DET cross section of $4\times10^{-12}$ cm$^2$ for $CFCl_3$ on $D_2O$ ice. The DET reactions of halogenated molecules can generate reactive radicals and halogen anions,[9] and most of the halogen anions are trapped at ice surfaces.[9-14, 16] Also, ionic reactions on ice surfaces are well known to convert $Cl^-/Br^-$ effectively into photoactive $X_2$, HOX and XONO (X=Cl, Br), as is demonstrated on aqueous sea-salt particle surfaces by Finlayson-Pitts and co-workers[17-19] and on condensed $CF_2Cl_2$ films by Hedhili et al.[20].

The above experimental findings have provided a sound physics and chemistry foundation for the cosmic-ray (CR) driven electron-induced reaction (CRE) mechanism for forming the polar ozone hole, which was proposed about two decades ago.[9, 12, 21] Moreover, substantial data from both laboratory measurements and atmospheric observations have been reviewed.[22-24] The CRE mechanism, which is schematically shown in Fig. 1, has therefore been established.

The ozone hole is expected to appear in the spring polar stratosphere annually since the 1980s. The hole over the Antarctic is much bigger than that over the Arctic due to the major differences between the Antarctic and Arctic polar vortices. Ozone depletion is closely related to surface reactions on ice in polar stratospheric clouds (PSCs).[25] Far more PSCs are present over the winter Antarctic as the Antarctic stratosphere is colder and the Antarctic polar vortex is more stable and persistent than the Arctic. But it is generally believed that most of the mechanisms at work in the Antarctic are also present in the Arctic. Despite the measured decreasing ozone-depleting substances (ODSs) in the troposphere since around 1994, however, there are still frequent 'surprising' observations unexpected from the photochemical models.[26-28] In 2020, for example, the Arctic $O_3$ hole set a biggest record,[29] while the Antarctic $O_3$ hole was the longest-lasting and one of the largest and deepest holes since the ozone layer monitoring began 40 years ago, according to the WMO.[30] Even more difficult to understand from the photochemical models is why minimum $O_3$ holes were observed in years of solar maxima (1991, 2002 and 2013), while largest, deepest and most persistent $O_3$ holes appeared in years of solar minima (1987, 1998, 2008 and 2020), with a regular periodicity of ~11 year over the past 4 solar cycles. Instead, these observations agree well with the prediction of the CRE model.[12, 21-24] Unfortunately the CRE mechanism has been completely ignored in current atmospheric chemistry context. This ignoring is most likely responsible for the persistent quantitative discrepancies between observations and (photo)chemistry-clime models (CCMs), as pointed out previously.[24] With new observed data from atmospheric measurements over the past decade, this Article will examine the key predictions of the CRE mechanism, provide fingerprint evidence of this mechanism and make new predictions on future observations.

**PREDICTIONS OF THE CRE MODEL VS PHOTOCHEMISTRY-CLIMATE MODELS**

In the Antarctic vortex, the measured ClO abundance of the order of 1 p.p.b.v. is two orders of magnitude higher than the usual concentration (0.01 p.p.b.v.) in the general stratosphere.[31, 32] This shows strong evidence of the $O_3$ hole being related to chlorine-containing molecules mainly CFCs. Furthermore, observations of large Antarctic $O_3$ holes in springs since the mid-1980s appear to support the photochemical mechanisms. Looking closely into observed data, however, one can see that there are actually large discrepancies between photochemical models and observations.[24] In the late 1990s, it was called attention to the significant gaps in understanding of the partitioning of chlorine in the stratosphere, as evidenced by persistent quantitative inconsistences between observed data and photochemical models.[33] For example, it was noted that modeled ClO/HCl concentration ratios in the lower stratosphere below 20 km were smaller than in situ measured values by a factor of at least two. Despite the passing of two decades, the situation remains essentially unchanged: today significant discrepancies between state-of-the-art CCMs and observations still exist, especially in the lower stratosphere and at midlatitudes.[34-36] In fact, the model predictions documented in a series of WMO Scientific Assessments of Ozone Depletion in 1994-2018 have kept changed but have rarely been proven. Although the signs of recovery in the Antarctic $O_3$ layer have been reported,[22-24, 34-38] the lower stratospheric $O_3$ layer at mid-latitudes has 'surprizingly' shown a continuous decreasing trend, in contrast to the observed steady declining of the effective chlorine in the troposphere since ~1994. This discrepancy is still dramatic.[23, 24, 34-36] Chemistry transport models (CTMs) require the use of observed temperatures and winds to produce the desirable outputs. This requirement not only limits CTMs' ability to predict future changes of the $O_3$ hole[39] but leads to their intrinsic problems as stratospheric cooling (temperature drop) is a well-known direct consequent result of $O_3$ loss ($O_3$ itself is a greenhouse gas)[33, 40, 41]. Indeed, there exists a nearly perfect linear correlation between observed total $O_3$ and temperature in the Antarctic $O_3$ hole with an obtained correlation coefficient of up to 0.94.[22-24] Thus, any CTM or CCM using historical meteorology such as measured stratospheric temperatures and polar vortex dynamics as model input parameters actually uses the $O_3$ depletion results to get the "right" output for $O_3$ depletion.

Below, the key observations and predictions of the CRE mechanism are briefly reviewed. First, a strong spatial correlation between CRs and $O_3$ loss exists in the stratosphere,[12] in which electrons are produced mainly from CR-initiated atmospheric ionization. The CRE reaction should show strong latitude and altitude effects, corresponding to the distribution of CR-produced electrons. Due to the geomagnetic effect, the intensity of CRs composed mainly of charged particles varies strongly with latitude, showing maxima at polar regions. Indeed, most severe $O_3$ loss has been observed at the polar stratosphere, while only minimum $O_3$ loss has been seen in the tropic since the 1970s. Moreover, the ionization rate of CRs depends strongly on altitude, exhibiting a maximum at 15−18 km from the ground, the so-called Pfotzer maximum. The observed data on $O_3$ loss over both mid-latitudes and springtime polar regions have shown two maxima: one is peaked at ~15 km in the lower stratosphere and the other at ~40 km in the upper stratosphere.[31, 42] The $O_3$ loss peak at ~40 km is consistent with the photodissociation mechanism.[26-28] However, the well-known (springtime) $O_3$ hole is exactly peaked at 15−18 km in the lower polar stratosphere. This study will focus on the mechanism for $O_3$ depletion in the lower stratosphere below 25 km, in which current discrepancies between CCMs and observations are most significant.[23, 24, 34-36]

Second, there also exists a strong temporal correlation between CRs and $O_3$ depletion. The CRE model predicted an ~11-year cyclic variation of $O_3$ loss in the polar $O_3$ hole, corresponding to the cyclic variation of the CR intensity that is anti-phased with the solar intensity with an average periodicity of ~11 years.[12, 22-24] A time correlation between the annual mean total $O_3$ in the southern hemisphere (at latitudes 0–65º S) and the CR intensity in the single CR cycle of 1979-1992 was first reported by Lu and Sanche in 2001.[12] It was immediately criticized that no such 11-year cyclic variations could exist beyond one CR cycles,[43] while we gave a rebuttal to such a criticism.[44] Subsequently, the author showed that there indeed existed 11-year cyclic CR-$O_3$ time correlations not only at near global (60º N–60º S) or semi-global (0–65º S) but in the springtime Antarctic ozone hole (60-90º S) over the two CR cycles up to 2008 or 2013.[21-24] Müller and Grooß[45-48] had repeated arguments for no correlation between CRs and $O_3$ loss in the polar region. As revealed previously, however, either the "satellite data" used in their papers were problematic[49, 50] or their criticisms cannot stand from the examination by reliable data and other scientific facts[24]. It is well-known that meteorological conditions such as atmospheric dynamics and volcanic eruptions can cause large fluctuations of the polar ozone hole from year to year. However, the author has found that a minimal processing by 3-year smoothing (adjacent averaging) of measured data can effectively reduce the fluctuation level and minimize the unpredictable short-term effects.[22-24] After such minimal processing, pronounced 11-year cyclic oscillations in total $O_3$ in the Antarctic ozone hole became manifest.[21-24] This minimal 3-year average processing has also been adopted recently by others.[38]

Furthermore, a concise equation was derived from the CRE mechanism, giving the percent change of total $O_3$ in the polar ozone hole by:

$$\Delta[O_3] = -k\,[C]I^2 \qquad (1)$$

where $[C]$ is the 'equivalent effective chlorine (EECl)' in the polar stratosphere, $I$ is the CR intensity, and $k$ is a fitting constant determined by the best fit to past measured $O_3$ data.[22-24] The quadratic dependence of polar $O_3$ loss on the CR intensity $I$ arises from the fact that the CRE reaction efficacy ($\chi$) is linearly proportional to the product of the electron density ($n_e$) produced by CRs and the coverage ($\theta_{halo}$) of charge-induced adsorption of halogenated molecules on the surface of ice particles in PSCs, where $n_e$ is linearly dependent on the CR intensity $I$ and $\theta_{halo}$ is also a linear product of $[C]$ and $I$.[24] That is, $\chi \propto n_e \theta_{halo}$, where $n_e \propto I$ and $\theta_{halo} \propto [C]I$, so that $\chi \propto [C]I^2$. It was once debated whether the surface of PSC ice would be sticky to halogen gases at the atmospheric temperature.[51, 52] However, it has been experimentally demonstrated that the CR-produced charge in PSCs can indeed induce effective adsorption of halogenated molecules including CFCs on the surface of PSC ice.[24] The sunlight plays an important role in destroying $O_3$ in the spring polar stratosphere (see Fig. 1), but there are abundant solar photons for the photolysis of halogen species. Hence, the solar intensity is not a limiting factor in $O_3$-depleting reactions and is not expressed in Eq. 1. For details on Eq. 1, the readers are referred to ref.[24].

Third, a similar 11-year cyclic variation of polar stratospheric cooling was also observed. $O_3$ loss is well-known to cause a stratospheric cooling.[40, 41] Thus, temperature drop in the lower polar stratosphere is a direct indicator of $O_3$ loss. The non-monotonic decreases in lower stratospheric temperature during 1979-1990 and 1979-2003 were once attributed to the effect of volcanic eruptions[40, 41]. However, it has now been well demonstrated that the lower polar stratospheric temperature regulated by stratospheric cooling due to $O_3$ loss has pronounced 11-year cyclic variations and can be well reproduced by Eq. 1.[22-24] Moreover, it is important to note that there have been no such 11-year oscillations observed in time-series data of the lower stratospheric temperature averaged in months of winter (May-August) prior to the appearance of the large $O_3$ hole in spring (September-December) over Antarctica. This fact has ruled out the direct effect of CRs without CFCs for the observed 11-year cyclic variations of polar stratospheric cooling.[22-24]

Last but not least, the CRE model distinguishes from photochemical models in the following prediction. The photochemical models assume that CFCs would mainly decompose by solar photolysis in the upper tropical stratosphere; air carrying the photoproducts (inorganic species) is then transported to the lower Antarctic stratosphere, where their heterogeneous chemical reactions on ice surfaces in PSCs occur to produce photoactive halogens in winter. Thus, the stratospheric age of $O_3$-depleting halogenated gases at mid-latitudes is younger at ~3 years, versus ~6 years over Antarctica. As a result, the recovery of the ozone layer at mid-latitudes or near global to the 1980 level would precede that over Antarctica by 10-20 years. In contrast, the CRE model gives that the *in-situ* CRE reaction of halogenated molecules including organic and inorganic molecules (CFCs,

HCl, ClONO$_2$, *etc.*) adsorbed or trapped at the surface of PSC ice is the key and limiting step to form (photo)active halogen species, leading to O$_3$ loss in the polar stratosphere in both winter and springtime (see Fig. 1). At mid-latitudes, the CRE reaction of halogenated molecules is far less effective due to the much weaker CR intensity and the lack of PSCs. Therefore the stratospheric EECl at Antarctica should be more sensitive to the tropospheric halogen change than that at mid-latitudes. This led to the CRE prediction that the recovery of the Antarctic O$_3$ hole is more closely tracking the declining halogen loading measured in the troposphere with a delay time of 1-2 years only, whereas the O$_3$ recovery at mid-latitudes occurs much later with a delay time of ≥10 years.[23,24] Both are in good agreements with recent observations.[34-36]

**NEW OBSERVED RESULTS**

This article will first examine closely the predictions made by the CRE model and CCMs with updated observed data up to 2020. Figs. 2A-C present both measured and projected time-series annual mean CR intensity, equivalent effective chlorine in the troposphere (tropospheric EECl, without transport and mixing lag times considered), October monthly mean Antarctic (60°-90°S) and annual mean near-global (60°S-60°N) total O$_3$ data observed by NASA satellites and given by the CRE equation with delay times of 1 year and 10 years for stratospheric EECl from tropospheric EECl at Antarctica and mid-latitudes respectively and the multi-model mean (MMM) from the projections of multiple CCMs[53]; Figs. 2D-F show observed time- series data of 3-month (October-December) mean total O$_3$ over Antarctica (60-90° S), 2-month (October-November) mean total O$_3$ over Syowa (69°S, 39.6°E), Antarctica and 3-month (October-December) mean total O$_3$ over South Pole (90°S) during 1979-2020 and those given by Eq. 1, respectively. Fig. 2A shows that the measured halogen loading in the troposphere had a significant rise from the late 1970s to 1994~1995, and since then a decreasing trend; whereas the measured CR intensity has exhibited a rising trend over the past 4 solar cycles, in good agreement with one of our projections.[22-24] During 1979-1995, there were corresponding drops in total O$_3$ over Antarctica and mid-latitudes (near global), as shown in Figs. 2B-F. After 1995, the O$_3$ layers at both Antarctica and near global have shown complicated trends, which require delicate analyses. First, minima (maxima) in ozone loss were indeed observed in 1991(1998), 2002 (2008) and 2013 (2020), corresponding to minima (maxima) in CR intensity over the past 3-4 solar cycles (most visible in Fig. 2E). There were unexpected large and small Antarctic ozone holes in 2015 and 2019 respectively, but they were due to the volcanic eruption of Calbuco in 2015 and the unusually stratospheric warming over Antarctica in early spring of 2019 (a rare event in the last 30 years) respectively.[37, 53] Both unexpected effects were most significant near tropopause but decreasing with rising altitudes (15-25 km) in the stratosphere, as revealed by observed data shown later. Second, both Antarctic (Figs. 2B, D-F) and near-global total O$_3$ (Fig. 2C) exhibit clear 11-year cyclic variations, as predicted by the CRE mechanism. Indeed, Eq. 1 has given excellent agreements with all the observed data from 1979 to 2020. Third, the CRE model predicted that the Antarctic O$_3$ will return to its level at 1980 by 2055-2060 (Fig. 2B),[23] while the mid-latitudes or near-global will not even by the end of this century (Fig. 2C), if the CR intensity keeps the rising trend observed in the past 4 solar cycles (as projected in Fig. 2A).[22-24] In contrast, the MMM of multiple CCMs, in spite of their use of historical meteorology such as stratospheric temperatures and polar vortex dynamics to account for atmospheric dynamical variability, has large discrepancies from the observed data. These inconsistences are especially notable for the period since 1995 and for ozone at mid-latitudes or near-global. In the 2018 WMO/UNEP Report[53], the CCMs have revised their projections on the O$_3$ recovery to the 1980 levels by 2055-2060 for the Antarctic and by 2040-2045 for near-global, as shown in Figs. 2B and C, which are about 10 years faster than those given in the 2014 WMO/UNEP Report. Although the revised recovery date for Antarctic O$_3$ by CCMs is now nearly the same as that given by the CRE model,[23] there still exist large gaps between CCMs and observations, especially for O$_3$ at mid-latitudes or near global. These discrepancies, most significant for the ozone trend in the lower stratosphere, were also reported in recent literature.[34-36] They must be resolved.

Figs. 3A-B show observed time-series data of 3-month (October-December) mean lower stratospheric temperature at 100 hPa (~15 km) over Syowa (69°S, 39.6°E) and South Pole (90°S), Antarctica during 1979-2020 and those given by Eq. 1, respectively. Again, it is clearly shown that stratospheric cooling (temperature drop) caused by O$_3$ loss in the Antarctic O$_3$ hole exhibits clear 11-year cyclic variations, as predicted by the CRE mechanism.

Furthermore, to give quantitative and statistical analyses of observed data in terms of the CRE mechanism, both measured and 3-year smoothed 3-month (October-December) mean total O$_3$ in Antarctica (60-90° S) and lower stratospheric temperature data at Syowa station are respectively plotted versus the product of $[C]I^2$ in Figs. 4A and B, where linear fits gave rise to statistical correlation coefficients -R of up to 0.87 for 3-year smoothed O$_3$ and temperature data. These values are very close to those (0.90-0.91) obtained for similar total O$_3$ and lower stratospheric temperature datasets from Halley station (75°S, 26°W) observed up to 2013.[23, 24] The slightly lower correlation coefficients obtained in current analyses of observed datasets up to 2020 are due to the inclusion of the data of 2015 and 2019 with the above-mentioned abnormal events. These results have clearly demonstrated that both total O$_3$ and lower stratospheric temperature in the Antarctic O$_3$ hole are well described by the CRE mechanism and can be calculated by Eq. 1 with the fitting parameter $k$ determined from past observed data.

Importantly, new fingerprints of the CRE mechanism will further be unraveled with measured altitude profiles of O$_3$ and tropospheric-stratospheric temperatures at Syowa (69°S, 39.6°E), Antarctica, obtained from the Ozonesonde and Umkehr datasets of the JMA's Antarctic Meteorological Data.[54] These are shown in Figs. 5 and 6. Fig. 5A plots O$_3$ altitude profiles of the October monthly mean in the pre-ozone hole period of 1968-1980 and the ozone-hole period of 1991-1997,[55] deepest Antarctic O$_3$ holes observed in 1998, 2006, 2008 and 2020, as well as the attitude profile of the CR ionization rate. First, the measured total O$_3$ on 3rd October 2020 indeed set a lowest record of 128 DU at Syowa since the 1960s. Second, of particular interest is that the O$_3$ layer at 13.5-17.5 km, where the peak of the CR ionization rate is located, has constantly been completely depleted in the largest and deepest O$_3$ holes typically observed around the years of 11-year cyclic CR peaks. In this lower stratospheric layer of ~4 km, the CRE reactions caused drastic O$_3$ loss from the late 1970s to the early 1990s, but they have over-killed O$_3$ molecules and therefore the O$_3$ layer has been insensitive to the CR maxima in 11-year cycles since the late 1990s. In contrast, at the stratospheric layer of 17.5-21.5 km, corresponding to the upper stratosphere of the undisturbed O$_3$ profile peak, the column O$_3$ has only been partially depleted from the 1970s up to date. At this upper O$_3$ layer, the CR ionization rate is significantly dropping from its altitude peak, whereas O$_3$ is at the upper half of its maximum peak. Thus, O$_3$ loss at this layer is sensitive to the cyclic CR maxima. Third, Fig. 5B plots the altitude-profile O$_3$ loss of October monthly mean in the O$_3$-hole period of 1991-1997 with respect to the pre-O$_3$ hole period of 1968-1980 versus the square of the CR ionization rate ($I^2$); a linear fit to the data gives a correlation coefficient of 0.96, which is nearly perfect, given the measurement uncertainties in ozone and CR intensity. Consistent with the above-observed over-killing effect, the computer-given linear line is above the data curve at the lower polar stratosphere and is below the data curve at the upper polar stratosphere. Thus, this contrast profile difference of $O_3$ depletion at the lower and upper layers of stratospheric $O_3$ peak is precisely consistent with what is expected from the CRE mechanism and provides fingerprint evidence of the mechanism.

To examine the above conclusion critically, measured time-series October monthly mean vertical distribution partial column $O_3$ at various tropospheric-stratospheric layers (253-127, 127-63, 63-32, 16-8 and 2-1 hPa) and October monthly mean atmospheric temperatures at 500, 300, 200, 150, 100, 50 and 30 hPa at Syowa (69°S, 39.6°E) as well as those given by the CRE equation during 1970-2020 are respectively shown in Figs. 6A and B. These altitude distribution data provide rich and valuable information on $O_3$ depletion mechanisms. First, the attribution of the dips in both $O_3$ and stratospheric temperature in 2015 to the volcanic eruption of Calbuco is well confirmed as they were most visible at tropopause, became more insignificant with increasing altitudes in the stratosphere and finally almost disappeared in the upper stratosphere of ≤30 hPa. A similar behavior of the ozone and temperature peaks in 2019 due to an unusually stratospheric warming was also observed, which became weaker at higher altitudes of ≤30 hPa. Second, partial column ozone at both stratospheric layers of 127-63 hPa and 63-32 hPa made the lowest records in 2020 in observation history since the 1960s, consistent with the appearance of the strongest CR peak in the current solar cycle. The October mean stratospheric temperatures at 100 hPa and 50 hPa in 2020 exhibited second lowest records only, slightly higher than those of 2006. However, this inconsistency with the lowest column ozone records was probably due to the measurement uncertainties in temperatures as the 3-month (October-December) mean stratospheric temperatures did clearly show the lowest records in 2020 (see Figs. 3 and 6B). Third, most remarkably, both column ozone at 63-32 hPa and stratospheric temperatures at 50 hPa have shown the best agreements with the CRE model, whereas their respective data at 127-63 hPa and 100 hPa have exhibited lower response sensitivities to the CR-driven cyclic variations than those given by the CRE model. This is consistent with the observed data shown in Fig. 5 and is due to the over-killing effect of the CR-driven reaction at the CR-peak polar stratospheric region. Fourth, with decreasing CR ionization rates at the troposphere (≥200 hPa) and upper stratosphere (16-8 and 2-1 hPa), the 11-year cyclic variations caused by the CRE effect diminish, as seen in both partial column $O_3$ and temperature data. It is particularly interesting to note that the time-series observed data of partial column $O_3$ in the upper polar stratosphere at 2-1 hPa (~40 km), where the photochemical mechanism dominates, exhibit no 11-year cyclic variations. This is consistent with the time series observed upper stratospheric annual mean $O_3$ anomalies at 2 hPa from multiple datasets by Chipperfield et al.[56]. Similarly, 11-year cyclic variations for the lower stratosphere at 50 hPa but not for the upper stratosphere at 5 hPa could also be found from the NASA's the GOZCARDS merged satellite dataset[57] (see Figs. S10 and S11 of ref.[58]), whereas no 11-year cyclic variations of partial column $O_3$ for both lower and upper polar stratospheres were shown from the Bodeker Scientific Vertical Ozone (BSVertOzone) database (see Figs. 5 and 6 of ref.[58]). However, the BSVertOzone dataset was not pure observations but was corrected 'using output from a chemical transport model' (cited from the database website) and 'optimised for use in comparisons with CCM simulations.[58] Overall, the time-series data of altitude distribution $O_3$ and temperature shown in Figs. 6A and B are excellently consistent with the vertical profiles of $O_3$ shown in Fig. 5. Together, these observed data provide convincing fingerprints of the CRE mechanism.

The data presented in Figs. 2-6 have robustly demonstrated that both $O_3$ loss and stratospheric temperatures in the Antarctic $O_3$ hole over the past 4 solar cycles are well reproduced by Eq. 1 that produces $O_3$ loss with the inputs of two variables only, i.e., the stratospheric EECl level [$C$] and CR intensity $I$. The CR intensities have been well recorded since 1960s, showing a rising trend in the past four solar cycles, which excellently agrees with one of the projections made in the previous studies.[22-24] This means that the signs in recovery of recorded mid-latitude (near-global) and polar $O_3$ losses are complicated by the mixed effect of rising CR intensities and declining EECl levels in the stratosphere. Thus, the real effectiveness of the Montreal Protocol, i.e., the change of the stratospheric EECl levels, can be unraveled by correcting measured stratospheric $O_3$ or temperature data with the CR-factor of $1/I^2$.[22-24] Here the corrected total $O_3$ data at the Antarctic and near-global from observed data updated to 2020 are shown in Figs. 7A and B, respectively, in which polynomial fits to the data give $R^2=0.71$ and $0.79$ (coefficient of determination) with the probability $P<0.0001$ for $R^2 = 0$ (no trend). Fig. 7A shows that corrected $O_3$ loss in the springtime Antarctic hole at 60-90° S has had a clear and steady recovery since around 1995. Comparing the $O_3$ data in Fig. 7A with the EECl data in Fig. 2A, one can clearly confirm that the corrected spring $O_3$ loss over Antarctica has closely followed the tropospheric measured EECl, with a short delay of only ~1 year in the polar stratosphere. In contrast, Fig. 7B shows that the recovery of $O_3$ loss at near-global (60°S-60°N) has a time delay of ≥10 years from the peak of the tropospheric measured EECl, clearly confirming that the stratospheric EECl decline and associated $O_3$ recovery in near global are significantly delayed, compared with those in the Antarctic stratosphere. This is also consistent with the findings in recent literature.[34-36] It is worth noting that the simple correction by the CR factor has indeed been validated by time series measured total ozone data (without correction by the CR factor) in the summertime stratosphere at both Antarctica and Halley station (75.4°S, 26.4°W) when there were no or few PSC ice particles. The latter data have directly exhibited the recovery of the Antarctic ozone layer since 1995, as shown in Figs. 8A and B with observed data updated to 2020. In the latter case, the change of measured total ozone directly reflects the change of the Antarctic stratospheric EECl. These observations, reported previously[22-24] and re-presented in current Figs. 7 and 8 including observed data updated to 2020, have provided solid evidence of the recovery of the Antarctic ozone hole. Recently, similar signs of recovery in the Antarctic $O_3$ hole were also reported by Solomon et al.[37] showing measured $O_3$ profile trends in September for the Antarctic stations and by Banerjee et al.[38] showing the changes in the summertime jet stream in the Southern Hemisphere. Overall, the results shown in Figs. 7 and 8 and those presented previously[22-24] not only provide strong evidence of the CRE mechanism but indicate that the Montreal Protocol has been successfully executed, placing the Montreal Protocol on a firm scientific ground.

**DISCUSSION**

The CRE equation (Eq. 1) has the single fitting parameter $k$ and it is challenging to derive this constant theoretically as CRs affect not only the formation of electrons but the charge-induced adsorption of halogenated molecules on the surface of PSC ice particles.[24] Nevertheless, it is worthwhile to make a quantitative estimate of $O_3$ loss by the CRE mechanism in the stratospheric region near the CR peak. The rate for production of electrons by CR ionization is ~45 cm$^{-3}$ s$^{-1}$ at an altitude of ~15 km in the general (gaseous) stratosphere. In the winter and early spring polar stratosphere, however, the situation is drastically different due to the presence of PSCs.[9, 12, 24] First, the CR ionization rate is four orders of magnitude higher, being ~2.3x10$^5$ cm$^{-3}$ s$^{-1}$ on the surface of ice particles in PSCs.[12] Second, the CR-produced charges in PSC ice particles can greatly enhance the adsorption of halogenated molecules on the surface, as is well demonstrated experimentally.[24] Third, since the DET reactions of ozone-depleting chlorine- and bromine-containing molecules adsorbed on ice surfaces are extremely effective, the CR-produced electrons in PSCs will rapidly be converted into mainly Cl⁻ (and minorly Br⁻) ions, most of which are stably trapped at PSC ice surfaces by the image potential.[9-14, 16] Given a lifetime of 1000 s for Cl⁻ on the surface of ice particles in the polar stratosphere,[59] the Cl⁻ density in PSC is expected to have an estimated equivalent volume density of $2.3 \times 10^8$ cm$^{-3}$ in the winter polar stratosphere. The known reactions on ice surfaces can effectively convert Cl⁻/Br⁻ into photoactive $X_2$, HOX and XONO (X=Cl, Br),[9, 24] as demonstrated experimentally.[17-20] It is worth noting that since there are abundant photons from the sunlight in the spring polar stratosphere, the solar intensity is not a limiting factor in ozone-depleting reactions.[24] This is evidenced by the observed facts of largest and deepest ozone holes in years of solar minima during 11-year cycles and of absent 11-year cyclic variations for ozone in the upper stratospheric at ~40 km. Assuming that 20% of the Cl⁻ ions are eventually converted into Cl atoms, the Cl atoms generated from CRE reactions in PSCs around the CR-peak polar stratospheric region are estimated to be $4.6 \times 10^7$ cm$^{-3}$. Given the well-known efficiency of one Cl atom being able to destroy $10^5$ $O_3$ molecules on average via well-known $O_3$-destroying reaction cycles,[26-28] these Cl atoms formed by the CRE mechanism could completely deplete the $O_3$ layer which typically has an $O_3$ density of $(4-5) \times 10^{12}$ cm$^{-3}$ at the period prior to the $O_3$-hole season. This estimate indicates that the CRE mechanism is able to completely destroy or even over-kill the $O_3$ layer in the CR-peak polar stratospheric region during CR maxima, in good agreement with the observations in Figs. 5 and 6.

Despite including the 'effects' of non-halogen greenhouse gases (GHGs) and allowing a number of metric choices (parameterisations) such as lower stratospheric temperature and its gradient and polar vortex dynamics to get the "right" outputs,[60] state-of-the-art CCMs still fail to reproduce the 11-year cyclic variations of stratospheric $O_3$ loss and associated stratospheric cooling and lead to large discrepancies with observations, especially at mid-latitudes and in the lower stratosphere.[34-36] As noted above, the temperature drop in the springtime $O_3$ hole is a direct consequence of $O_3$ loss.[22-24, 40, 41] Thus, any CCM or CTM using lower-stratospheric temperature and/or its profile in the $O_3$ hole as an 'input' to simulate the $O_3$ loss is seriously problematic in methodology. Such modeling is essentially a pseudo-calculation, using an input parameter (temperature) that is highly linearly correlated with an output (column ozone) to produce the output (column ozone). Notably, the prediction by CCMs that the recovery of the $O_3$ layer at mid-latitudes or near global would precede that over polar regions is just opposite to the observations. These discrepancies have been attributed to such factors as enhanced tropical upwelling, increased horizontal mixing and rising GHGs.[35] However, the results of Figs. 2-8 have robustly demonstrated that the observed data of stratospheric $O_3$ and temperatures over Antarctica in the past 40 years have been well described by Eq. 1 with the stratospheric EECl and the CR intensity as only two variables, indicating no signs of non-halogen GHG effects on the stratospheric temperature and ozone of Antarctica. This is in striking contrast to the predictions of CCMs.

## CONCLUSIONS

In summary, the proposed CRE mechanism of the $O_3$ hole has been convincingly confirmed by the observed data of stratospheric $O_3$ and temperatures over the past 40 years, as evidenced by excellently reproducing the observed data by Eq. 1 with the stratosphere equivalent chlorine level [*C*] and the CR intensity *I* as only two variables. The CRE predictions of 11-year cyclic variations of the Antarctic $O_3$ hole and associated stratospheric cooling have also been well proven. In contrast, no such cyclic variations have been observed for the upper polar stratosphere where the photochemical mechanism is dominant. The measured altitude profiles of $O_3$ and temperatures in the Antarctic $O_3$ hole also show that the CRE reaction, rather than the photoactivation of halogen species in the gas phase (due to the abundance of solar photons), is the limiting factor for $O_3$ loss in the springtime polar stratosphere. The quantitative estimate indicates that the Cl atoms produced by the CRE mechanism have the capacity to completely deplete or even over-kill the $O_3$ layer in the CR-peak polar stratospheric region during CR maxima, consistent with the observed altitude profiles of severe Antarctic $O_3$ holes. After the removal of the natural CR effect, the hidden recovery in the Antarctic $O_3$ hole since ~1995 has been clearly discovered, while the recovery of $O_3$ loss at mid-latitudes has been delayed by ≥10 years. The polar $O_3$ hole has shown a quick response to the decline in total halogen burden in the troposphere since 1995. All together, these results have provided fingerprint evidence of the CRE mechanism. If the CR intensity keeps the trend observed in the past 4 solar cycles, the Antarctic $O_3$ hole will return to the 1980 level by around 2060, while the returning of the $O_3$ layer in near global (mid-latitudes) to the 1980 level will be largely delayed or will not even occur by the end of this century. This study further predicts another minimum $O_3$ hole to observe around 2025±1 and another maximum $O_3$ hole around 2031±1.

The photolysis mechanism of CFCs can well explain the observed $O_3$ data for the upper stratosphere, and the gas-phase photoactivation to produce halogen atoms is undoubtedly one of the important processes for the formation of the $O_3$ hole in the lower polar stratosphere in spring. However, the (photo)chemistry-climate models have led to significant discrepancies with the observed data for the lower stratosphere and have been unable to predict or explain the robustly observed 11-year cyclic variations of the Antarctic $O_3$ hole and associated stratospheric cooling. The data presented in this study strongly indicate that the CRE mechanism must be considered as a key factor in evaluating $O_3$ depletion in the lower stratosphere and the Antarctic/Arctic $O_3$ hole in order to remove the persistent discrepancies between CCMs and observations.

**ACKNOWLEDGEMENTS**

The author is greatly indebted to the above Science Teams for making these data available. This work is supported by the Canadian Institutes of Health Research and Natural Science and Engineering Research Council of Canada.

**AUTHOR DECLARATIONS**

The author has no conflicts to disclose.

**DATA AVAILABILITY STATEMENT**

The data that supports the findings of this study are openly available in the following sources: Total $O_3$ data over Antarctica and near-global (60°S-60°N) were obtained from NASA TOMS, OMI and OMPS satellite datasets. Altitude profiles of $O_3$, vertical partial column $O_3$ distribution at various atmospheric layers and total $O_3$ at Syowa were obtained from the WMO's WOUDC[61] and the Ozonesonde and Umkehr datasets of the Japan Meteorological Agency's (JMA's) Antarctic Meteorological Data[62]. Total ozone at South Pole (90°S) were obtained from the WOUDC[61] and the NOAA's Global Monitoring Laboratory[63]. Lower stratospheric temperature data at Syowa and Amundsen_Scott (South Pole) and total $O_3$ data at Halley were obtained from the British Antarctic Survey (BAS)[64]. Cosmic ray data were obtained from McMurdo station up to 2016[65] and by being relative to the measured data at Moscow since 2017[66], and projected CR data were obtained from references[22-24]. Tropospheric EECl data were from the NOAA's Global Monitoring Laboratory or from the 2018 WMO Report[53].

**REFERENCES:**

1. W. Hickam and D. Berg, Negative ion formation and electric breakdown in some halogenated gases. J. Chem. Phys. **29**, 517-523 (1958).
2. L. G. Christophorou, Electron-attachment to molecules in dense gases (quasi-liquids). Chem. Rev. **76**, 409-423 (1976).
3. E. Illenberger, H. Scheunemann, and H. Baumgartel, Negative-ion formation in $CF_2Cl_2$, $CF_3Cl$ and $CFCl_3$ following low-energy (0-10 ev) impact with near monoenergetic electrons. Chem. Phys. **37**, 21-31 (1979).
4. S. D. Peyerimhoff and R. Buenker, Potential curves for dissociative electron-attachment of $CFCl_3$. Chem. Phys. Lett. **65**, 434-439 (1979).
5. M. Lewerenz, B. Nestmann, P. Bruna, and S. D. Peyerimhoff, The electronic-spectrum, photodecomposition and dissociative electron-attachment of $CF_2Cl_2$ - an abinitio configuration-interaction study. J. Mol. Struct.-Theochem **24**, 329-342 (1985).
6. F. Fehsenfeld et al. Ion chemistry of chlorine compounds in troposphere and stratosphere. J. Geophys. Res.-Oceans Atmos. **81**, 4454-4460 (1976).
7. D. Smith and N. G. Adams, Elementary plasma reactions of environmental interest. Top. Curr. Chem. **89**, 1-43 (1980).
8. D. G. Torr, The Photochemistry of the Upper Atmosphere in J. S. Levine, Ed. The Photochemistry of atmospheres: Earth, the other planets, and comets (Academic Press, 1985) Chapt. 5, pp. 165-278.
9. Q.-B. Lu and T. E. Madey, Giant enhancement of electron-induced dissociation of chlorofluorocarbons coadsorbed with water or ammonia ices: Implications for atmospheric ozone depletion. J. Chem. Phys. **111**, 2861-2864 (1999).
10. Q.-B. Lu and T. E. Madey, Negative-ion enhancements in electron-stimulated desorption of $CF_2Cl_2$ coadsorbed with nonpolar and polar gases on Ru(0001). Phys. Rev. Lett. **82**, 4122-4125 (1999).
11. Q.-B. Lu and L. Sanche, Enhanced dissociative electron attachment to $CF_2Cl_2$ by transfer of electrons in precursors to the solvated state in water and ammonia ice. Phys. Rev. B **63**, 153403 (2001).
12. Q.-B. Lu and L. Sanche, Effects of cosmic rays on atmospheric chlorofluorocarbon dissociation and ozone depletion. Phys. Rev. Lett. **87**, 078501 (2001).
13. Q.-B. Lu and L. Sanche, Large enhancement in dissociative electron attachment to HCl adsorbed on $H_2O$ ice via transfer of presolvated electrons. J. Chem. Phys. **115**, 5711-5713 (2001).

14. S. Ryu, J. Chang, H. Kwon, and S. Kim, Dynamics of solvated electron transfer in thin ice film leading to a large enhancement in photodissociation of CFCl$_3$. J. Am. Chem. Soc. **128**, 3500-3501 (2006).
15. M. Bertin et al., Reactivity of water-electron complexes on crystalline ice surfaces. Faraday Discuss. **141**, 293-307 (2009).
16. J. Stähler, C. Gahl, and M. Wolf, Dynamics and reactivity of trapped electrons on supported ice crystallites. Acc. Chem. Res. **45**, 131-138 (2012).
17. K. Oum, M. Lakin, D. DeHaan, T. Brauers, and B. J. Finlayson-Pitts, Formation of molecular chlorine from the photolysis of ozone and aqueous sea-salt particles. Science **279**, 74-77 (1998).
18. K. Oum, M. Lakin, and B. J. Finlayson-Pitts, Bromine activation in the troposphere by the dark reaction of O$_3$ with seawater ice. Geophys. Res. Lett. **25**, 3923-3926 (1998).
19. E. M. Knipping et al., Experiments and Simulations of Ion-Enhanced Interfacial Chemistry on Aqueous NaCl Aerosols. Science **288**, 301-306 (2000).
20. M. N. Hedhili et al., Low-energy electron-induced processes in condensed CF$_2$Cl$_2$ films. J. Chem. Phys. **114**, 1844-1850 (2001).
21. Q.-B. Lu, Correlation between Cosmic Rays and Ozone Depletion. Phys. Rev. Lett. **102**, 118501 (2009).
22. Q.-B. Lu, Cosmic-ray-driven electron-induced reactions of halogenated molecules adsorbed on ice surfaces: Implications for atmospheric ozone depletion and global climate change. Phys. Rep. **487**, 141-167 (2010).
23. Q.-B. Lu, Cosmic-ray-driven reaction and greenhouse effect of halogenated molecules: culprits for atmospheric ozone depletion and global climate change. Intl. J. Mod. Phys. B **27**,1350073 (2013).
24. Q.-B. Lu, New Theories and Predictions on the Ozone Hole and Climate Change (*World Scientific*, 2015) 1-285.
25. O. Toon and R. Turco, Polar stratospheric clouds and ozone depletion. Sci. Am. **264**, 68-74 (1991).
26. R. Stolarski and R. Cicerone, Stratospheric chlorine - possible sink for ozone. Can. J. Chem. **52**, 1610-1615 (1974).
27. M. J. Molina and F. S. Rowland, Stratospheric sink for chlorofluoromethanes - chlorine atomic-catalysed destruction of ozone. Nature **249**, 810-812 (1974).
28. F. S. Rowland, Stratospheric ozone depletion by chlorofluorocarbons (Nobel Lecture). Angew. Chem. Intl. Ed. **35**, 1786-1798 (1996).
29. A. Witze, Rare ozone hole opens over the arctic - and it's big. Nature **580**, 18-19 (2020).
30. World Meteorological Organization, News, "Record-breaking 2020 ozone hole closes", Published on 6 January 2021, https://public.wmo.int/en/media/news/record-breaking-2020-ozone-hole-closes.
31. P. Solomon et al., High-concentrations of chlorine monoxide at low altitudes in the antarctic spring stratosphere - secular variation. Nature **328**, 411-413 (1987).
32. J. G. Anderson, D. Toohey, and W. Brune, Free-radicals within the Antarctic vortex - the role of CFCs in Antarctic ozone loss. Science **251**, 39-46 (1991).

33. J. J. Orlando and S. Schauffler, Halogen Compounds in G. Brasseur, J. J. Orlando, and G. S. Tyndall, Ed. Atmospheric Chemistry and Global Change (Oxford University Press, 1999) Chapt. 8, pp. 307.
34. W. T. Ball et al., Evidence for a continuous decline in lower stratospheric ozone offsetting ozone layer recovery. Atmos. Chem. Phys. **18**, 1379-1394 (2018).
35. W. T. Ball, G. Chiodo, M. Abalos, J. Alsing, and A. Stenke, Inconsistencies between chemistry-climate models and observed lower stratospheric ozone trends since 1998. Atmos. Chem. Phys. **20**, 9737-9752 (2020).
36. C. Zerefos et al., Representativeness of single lidar stations for zonally averaged ozone profiles, their trends and attribution to proxies. Atmos. Chem. Phys. **18**, 6427-6440 (2018).
37. S. Solomon et al., Emergence of healing in the Antarctic ozone layer. Science **353**, 269-274 (2016).
38. A. Banerjee, J. Fyfe, L. Polvani, D. Waugh, and K.-L. Chang, A pause in southern hemisphere circulation trends due to the Montreal Protocol. Nature **579**, 544-548 (2020).
39. G. Manney et al., Unprecedented Arctic ozone loss in 2011. Nature **478**, 469-475 (2011).
40. V. Ramaswamy, M. Schwarzkopf, and W. J. Randel, Fingerprint of ozone depletion in the spatial and temporal pattern of recent lower-stratospheric cooling. Nature **382**, 616-618 (1996).
41. V. Ramaswamy et al., Anthropogenic and natural influences in the evolution of lower stratospheric cooling. Science **311**, 1138-1141 (2006).
42. Randel, W. J. et al. Atmosphere - Trends in the vertical distribution of ozone. Science **285**, 1689-1692 (1999).

43. P. Patra and M. Santhanam, Comment on "Effects of cosmic rays on atmospheric chlorofluorocarbon dissociation and ozone depletion". Phys. Rev. Lett. **89**, 219803 (2002).
44. Q.-B. Lu and L. Sanche, Lu and Sanche Reply. Phys. Rev. Lett. **89**, 219804 (2002).
45. R. Müller and J. Grooß, Does Cosmic-Ray-Induced Heterogeneous Chemistry Influence Stratospheric Polar Ozone Loss? Phys. Rev. Lett. **103**, 228501 (2009).
46. J. Grooß and R. Müller, Do cosmic-ray-driven electron-induced reactions impact stratospheric ozone depletion and global climate change? Atmos. Environ. **45**, 3508-3514 (2011).
47. R. Müller and J. Grooß, Comment on "Cosmic-ray-driven reaction and greenhouse effect of halogenated molecules: Culprits for atmospheric ozone depletion and global climate change". Intl. J. Mod. Phys. B **28**, 14820013 (2014).
48. R. Müller and J. Grooß, A note on the 'Reply to 'Comment on "Cosmic-ray-driven reaction and greenhouse effect of halogenated molecules: Culprits for atmospheric ozone depletion and global climate change" by Rolf Müller and Jens-Uwe Grooß by Q.-B. Lu'. Intl. J. Mod. Phys. B **28**, 14750022 (2014).
49. Q.-B. Lu, Reply to "Comment on 'Cosmic-ray-driven reaction and greenhouse effect of halogenated molecules: Culprits for atmospheric ozone depletion and global climate change' by Rolf Müller and Jens-Uwe Grooß". Intl. J. Mod. Phys. B **28**, 14820025 (2014).
50. Q.-B. Lu, Author's reply to a Note on the Reply to Comment on "Cosmic-ray-driven reaction and greenhouse effect of halogenated molecules: Culprits for atmospheric ozone depletion and global climate change". Intl. J. Mod. Phys. B **28**, 14750034 (2014).

51. N. R. P. Harris, J. C. Farman and D. W. Fahey, Comment on "Effects of cosmic rays on atmospheric chlorofluorocarbon dissociation and ozone depletion". Phys. Rev. Lett. **89**, 219801 (2002).
52. Q.-B. Lu and L. Sanche, Lu and Sanche Reply. Phys. Rev. Lett. **89**, 219802 (2002).
53. WMO/UNEP Global Ozone Research and Monitoring Project—Report No. 58. SCIENTIFIC ASSESSMENT OF OZONE DEPLETION: 2018.
54. K. Miyagawa, I. Petropavlovskikh, R. D. Evans, C. Long, J. Wild, G. L. Manney and W. H. Daffer, Atmos. Chem. Phys. **14**, 3945-3968 (2014).
55. WMO/UNEP Global Ozone Research and Monitoring Project—Report No. 44. SCIENTIFIC ASSESSMENT OF OZONE DEPLETION: 1998.
56. M. Chipperfield et al., et al. Detecting recovery of the stratospheric ozone layer. Nature **549**, 211-218 (2017).
57. L. Froidevaux et al., Global ozone chemistry and related trace gas data records for the stratosphere (GOZCARDS): methodology and sample results with a focus on HCl, H$_2$O, and O$_3$. Atmos. Chem. Phys. **15**, 10471-10507 (2015).
58. S. Dhomse et al., Estimates of ozone return dates from Chemistry-Climate Model Initiative simulations. Atmos. Chem. Phys. **18**, 8409-8438 (2018).
59. A. A. Viggiano, In-situ mass-spectrometry and ion chemistry in the stratosphere and troposphere. Mass Spectrom. Rev. **12**, 115-137 (1993).
60. M. Amos et al., Projecting ozone hole recovery using an ensemble of chemistry-climate models weighted by model performance and independence. Atmos. Chem. Phys. **20**, 9961-9977 (2020).
61. https://woudc.org/data.php.
62. http://www.data.jma.go.jp/antarctic/datareport/06_ozone/ozone_e.html.
63. http://www.esrl.noaa.gov/gmd/dv/spo_oz.
64. https://legacy.bas.ac.uk/met/data.html.
65. http://neutronm.bartol.udel.edu.
66. http://cr0.izmiran.ru/mosc/.

# Figure Captions

**Fig. 1.** The cosmic-ray-driven electron-induced reaction (CRE) mechanism of the ozone hole:[9, 12, 21-24] Cosmic-ray (CR) driven electron-induced reactions of halogen-containing molecules (e.g., CFCs, HCl, and ClONO$_2$) in polar stratospheric clouds (PSCs) result in the formation of Cl, ClO, CF$_2$Cl$^\bullet$ or CFCl$_2^\bullet$, and Cl$^-$ ions. The Cl$^-$ ions can either be rapidly converted to reactive Cl atoms or react with other species to release photoactive Cl$_2$ and ClNO$_2$ in the winter polar stratosphere. The Cl, ClO, CF$_2$Cl$^\bullet$ or CFCl$_2^\bullet$ can destroy the O$_3$ layer in the polar stratosphere during winter; the photolysis of Cl$_2$, ClNO$_2$, ClOOCl can also enhance ozone destruction upon the return of sunlight in the springtime.

**Fig. 2**. A-C: Observed and projected time-series data of cosmic ray (CR) intensity, equivalent effective chlorine (EECl) data from tropospheric measurements, total ozone in October Antarctica (60-90° S) (open squares) and annual near-global (60° S-60° N) (open circles), as well as those given by the CRE model (Eq. 1)[23, 24] and the multi-model mean (MMM) of CCMs obtained from the 2018 WMO/UNEP Report[53] during 1979-2100. A: both tropospheric EECl and CR data are normalized to their respective values in 1980. B and C: the dot lines show the returns of the projected Antarctic and near-global ozone to their respective values in 1980 by the CRE and the MMM of CCMs. D-F: Observed and projected time-series data of 3-month (Oct-Dec) mean total ozone in Antarctica (60-90° S), 2-month (Oct-Nov) mean total ozone at Syowa (69°S, 39.6°E) station and 3-month (Oct-Dec) mean total ozone at South Pole (90°S) station, as well as the theoretical data given by the CRE equation during 1979-2020. In B-F, also shown are the 3-year smoothing (thick solid line in black) to the observed data.

**Fig. 3**. A and B: Observed and projected time-series data of 3-month (Oct-Dec) mean lower stratospheric temperatures at 100 hPa (~15 km) at Syowa (69°S, 39.6°E) and South Pole (90°S) stations, Antarctica, as well as the theoretical data given by the CRE equation during 1979-2020. Also shown are the 3-year smoothing to the observed data.

**Fig. 4**. A and B: Statistical analyses of 3-month (October-December) mean total ozone in Antarctica (60-90° S) and lower stratospheric temperatures at Syowa station in terms of the CRE mechanism during 1979-2020. Both original (open squares and circles) and 3-year smoothed (solid squares and circles) observed data are plotted as a function of the product of $[C]I^2$; linear fits to the data give linear correlation coefficients -R indicated, and P<0.0001 (for R=0).

**Fig. 5.** A. Measured altitude profiles of the CR ionization rate $I$, the October monthly mean ozone in the pre-ozone hole period of 1968-1980 and the ozone-hole period of 1991-1997 (1998 WMO/UNEP Report[57]), and the deepest Antarctic ozone holes observed in 1998, 2006, 2008 and 2020 at Syowa (69°S, 39.6°E), Antarctica. B. The altitude-profile O$_3$ loss of October monthly mean in the ozone-hole period of 1991-1997 with respect to the pre-ozone hole period of 1968-1980 is plotted versus the square of the CR ionization rate ($I^2$), and the red line is produced by a linear fit to the data, giving a nearly perfect correlation coefficient of 0.96; the arrows show the altitude increasing direction.

**Fig. 6.** A and B: Time-series data of October monthly mean vertical distribution partial column ozone at various tropospheric-stratospheric layers (253-127, 127-63, 63-32, 16-8 hPa and 2-1 hPa) and October monthly mean atmospheric temperatures at various altitudes of 500, 300, 200, 150, 100, 50 and 30 hPa at Syowa (69°S, 39.6°E) as well as the theoretical data given by the CRE equation during 1970-2020. In B, time-series 3-month (October-December) mean temperatures at 30 hPa (orange, top) are also shown, and the 3-year smoothed data for 200 and 300 hPa (bottom) are offset as indicated to plot together with the data for 150 hPa.

**Fig. 7**. A and B: Recoveries of the O$_3$ layer over Antarctica (60-90° S) versus near-global (60°S-60°N). Both time-series 3-month (October-December) mean total O$_3$ loss in the springtime Antarctic O$_3$ hole and annual mean O$_3$ loss at near global during 1979-2020 are corrected by the CR factor $1/I^2$ to remove the CR effect. A 3-year smoothing (red solid lines) was applied to the observed data (open squares and circles); polynomial fits to the corrected data gave coefficients of determination R$^2$ indicated and P<0.0001 for R$^2$=0 (no trend). Adapted and updated from Lu[23, 24].

**Fig. 8.** A and B: Observed time-series summertime (January-March mean) total ozone at both Antarctica (60-90° S) and Halley station (75°S, 26°W) during 1979-2020. A 3-year smoothing (red lines) was applied to the observed data (open squares and circles). A pronounced recovery in the summer polar ozone was unraveled. Adapted and updated from Lu[22-24].

**Figure 1**

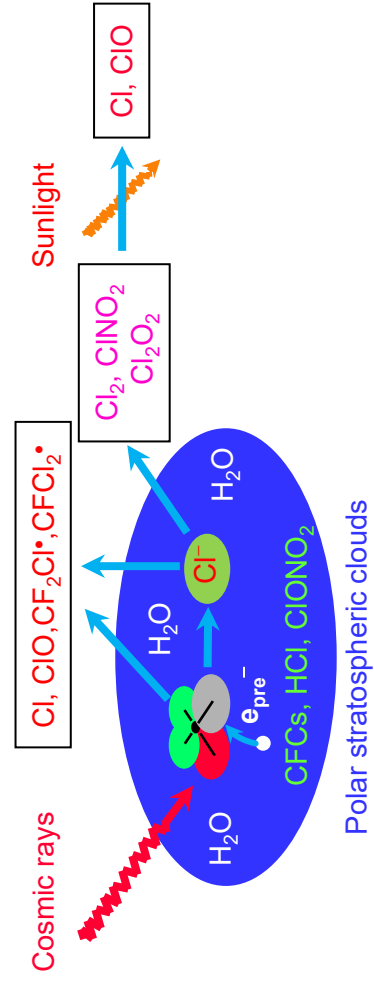

**Figure 2**

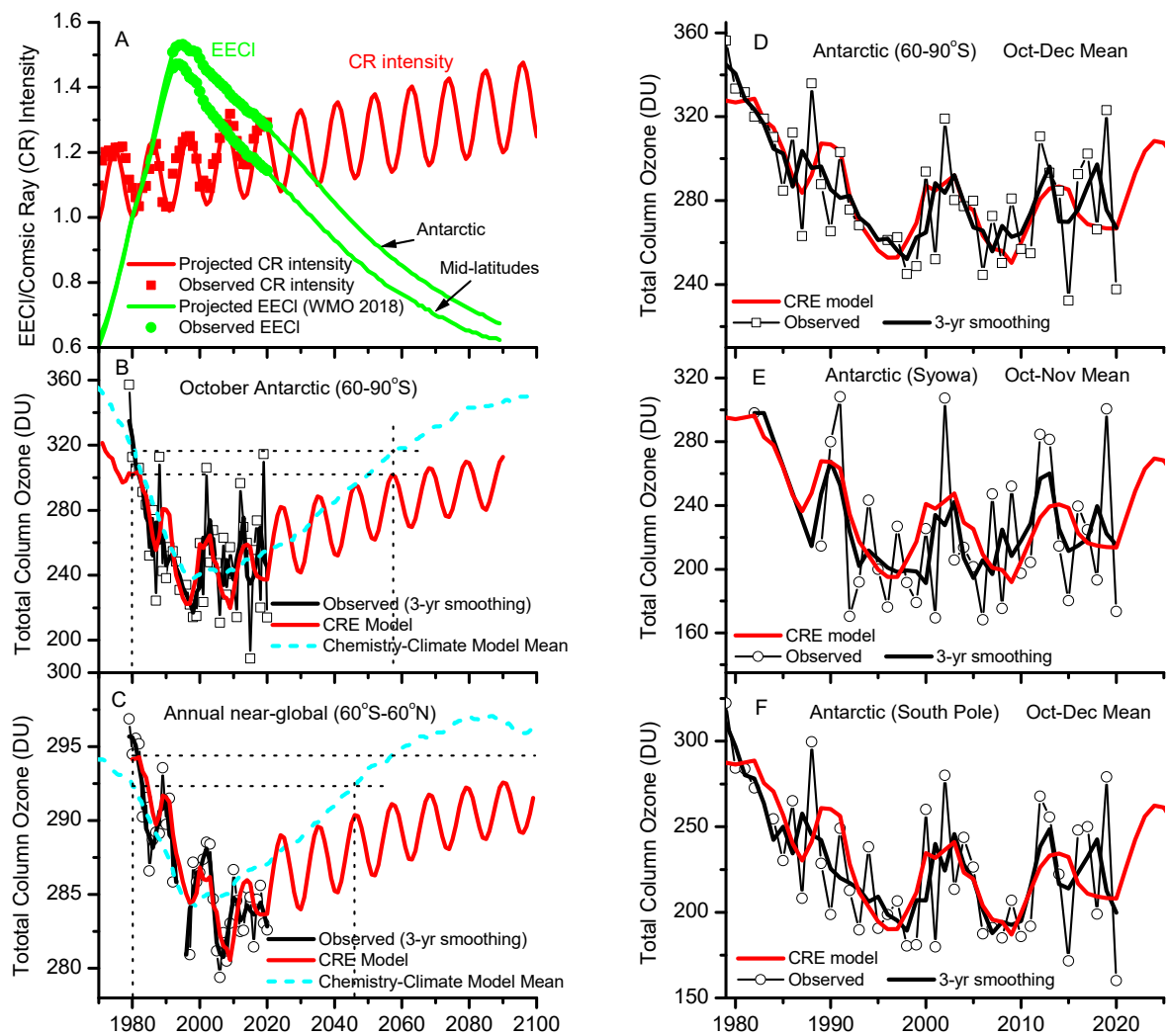

**Figure 3**

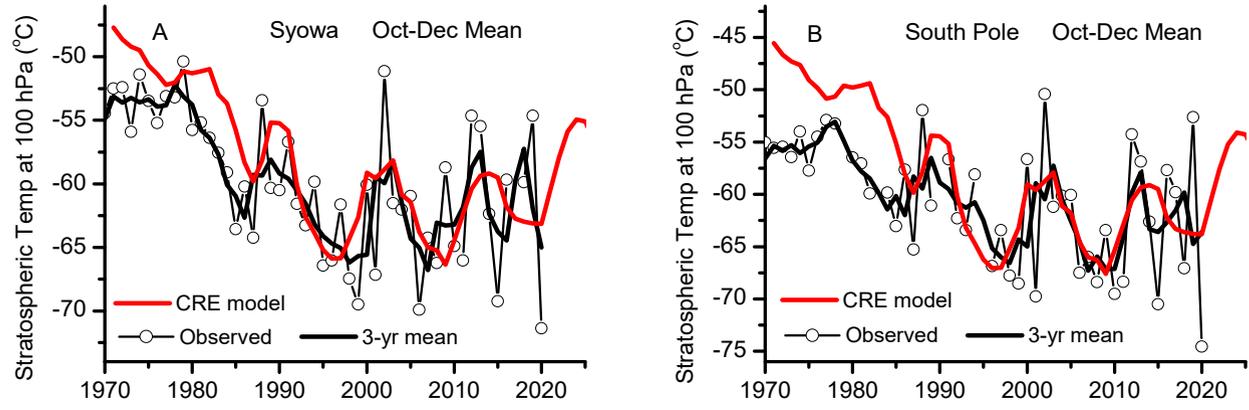

**Figure 4**

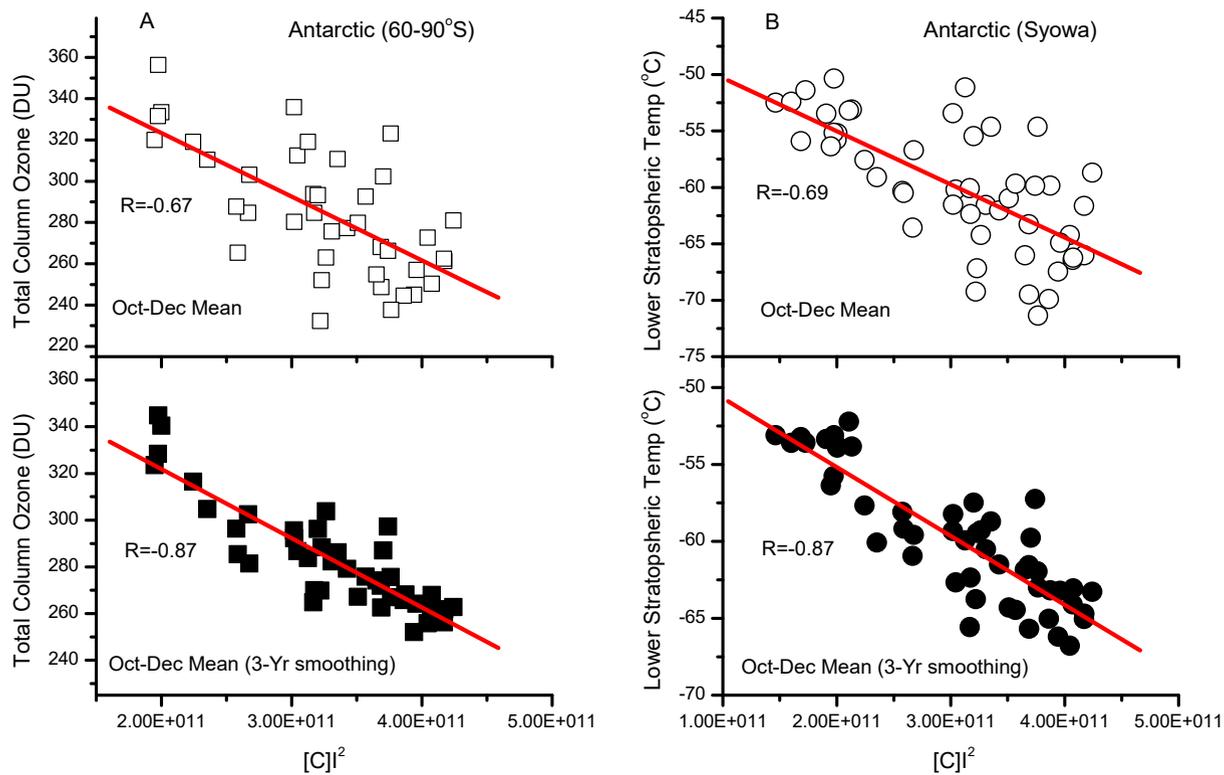

**Figure 5**

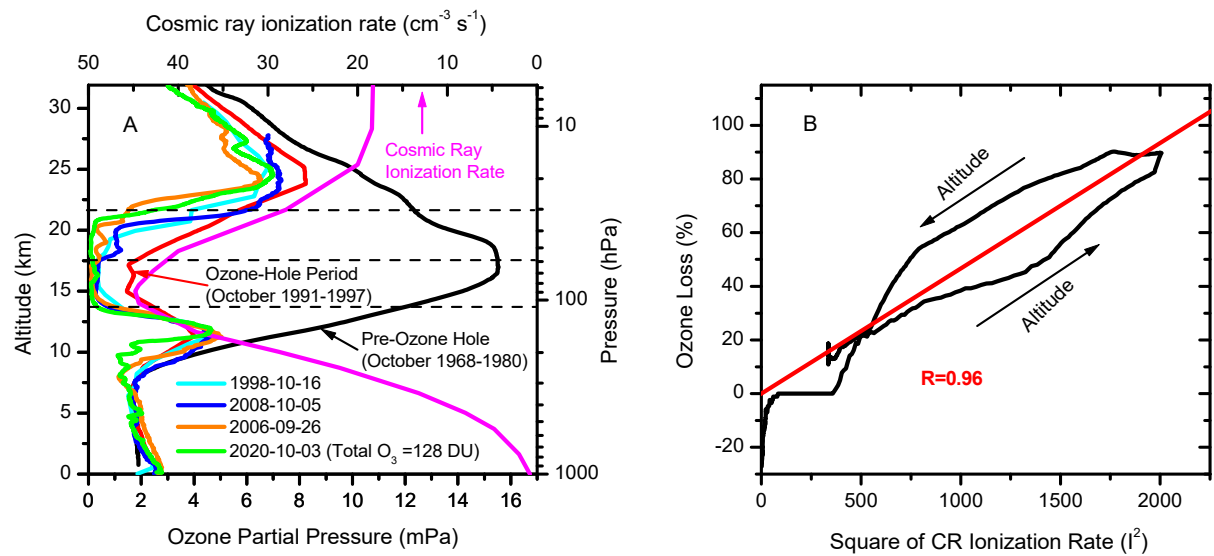

**Figure 6**

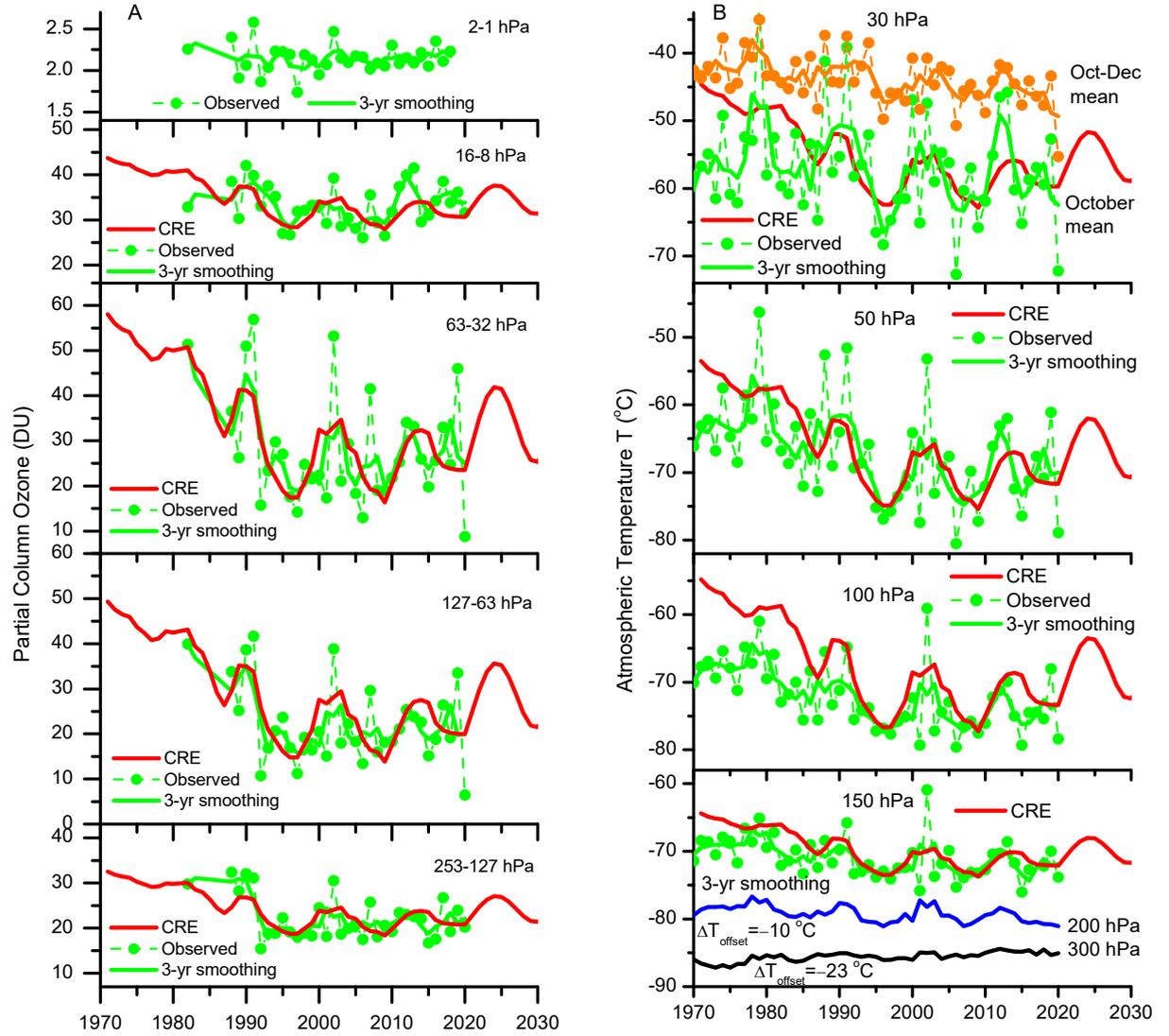

**Figure 7**

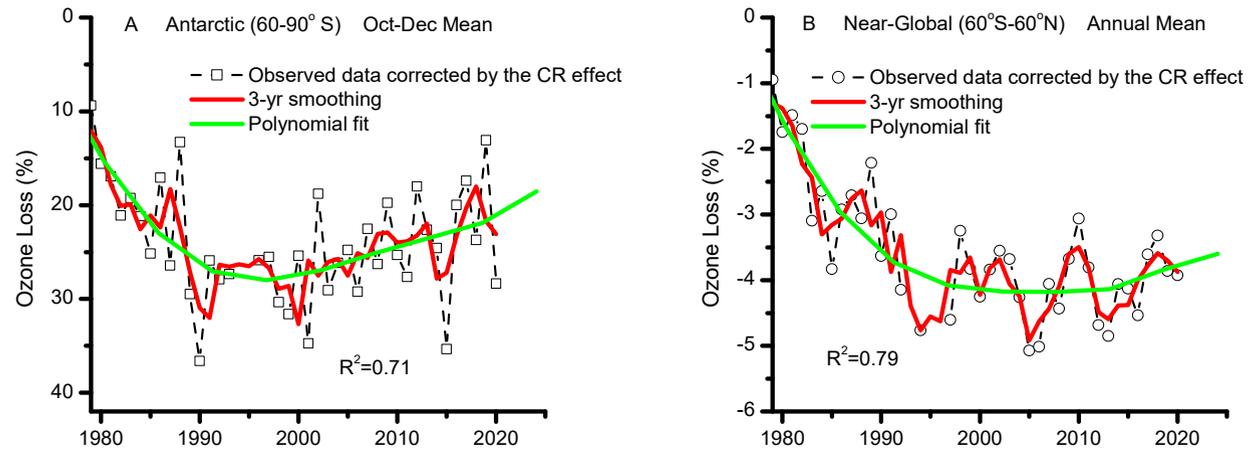

**Figure 8**

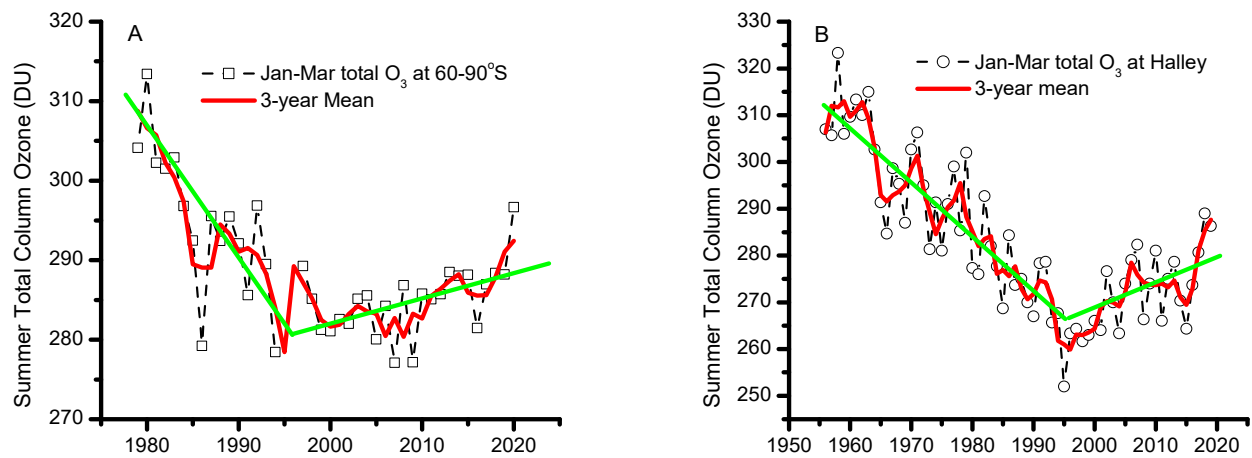